\documentclass[aps,showpacs,twocolumn,preprintnumbers,amsmath,amssymb]{revtex4}
\usepackage{amsmath,amssymb}
\usepackage{graphics,epsfig,mathtools}
\usepackage{graphicx}
\usepackage{color}
\def\build#1_#2^#3{\mathrel{\mathop{\kern 0pt#1}\limits_{#2}^{#3}}}

\renewcommand{\tilde}{\widetilde}          % wider `tilde'
\DeclareMathSymbol{\leqslant}{\mathalpha}{AMSa}{"36} % nicer `smaller or equal'
\DeclareMathSymbol{\geqslant}{\mathalpha}{AMSa}{"3E} % nicer `larger or equal'
\DeclareMathSymbol{\eset}{\mathalpha}{AMSb}{"3F}     % nicer `emptyset'
\def\build#1_#2^#3{\mathrel{\mathop{\kern 0pt#1}\limits_{#2}^{#3}}}

\def \E{ \mathbb E  }
\DeclarePairedDelimiter\floor{\lfloor}{\rfloor}

\begin{document}

\title{Regularized fractional Ornstein-Uhlenbeck processes, \\
and their relevance to the modeling of fluid turbulence}
%\\ . \\ \today}

\author{Laurent Chevillard$^{1}$}
\affiliation{$^1$Univ Lyon, Ens de Lyon, Univ Claude Bernard, CNRS, Laboratoire de Physique, 46 all\'ee d'Italie F-69342 Lyon, France}

\begin{abstract}
Motivated by the modeling of the temporal structure of the velocity field in a highly turbulent flow, we propose and study a linear stochastic differential equation that involves the ingredients of a Ornstein-Uhlenbeck process, supplemented by a fractional Gaussian noise, of parameter $H$, regularized over a (small) time scale $\epsilon>0$. A peculiar correlation between these two plays a key role in the establishment of the statistical properties of its solution. We show that this solution reaches a stationary regime, which marginals, including variance and increment variance, remain bounded when $\epsilon \to 0$. In particular, in this limit, for any $H\in ]0,1[$, we show that the increment variance behaves at small scales as the one of a fractional Brownian motion of same parameter $H$.  From the theoretical side, this approach appears especially well suited to deal with the (very) rough case $H<1/2$, including the boundary value $H=0$, and to design simple and efficient numerical simulations.
\end{abstract}


\maketitle

\section{Introduction}

The motivation of this present work comes from the stochastic modeling of certain aspects of fluid turbulence \cite{Fri95,Pop00}. In three dimensional space, a fluid, when stirred at large scale $L$ by a force, reacts while developing fluctuations: the fluid velocity $\textbf{u}(\textbf{x},t)$ reaches a statistically stationary regime characterized by a standard deviation, say $\sigma$. The more intense is the forcing, the bigger is $\sigma$. If it happens that the product $\sigma L$ is much bigger than the kinematic viscosity $\nu$ of the fluid, such that the so-called Reynolds number $\mathcal R_e=\sigma L/\nu$ is $\gg 1$, the fluid motions become turbulent. In this regime, velocity fluctuations, characterized by $\sigma$, are observed to be independent on the very nature of the dissipative mechanisms that are taking place, in particular of the precise value of $\nu$.  In order to achieve such a efficient way to dissipate energy, the fluid will develop small spatial scales such that asymptotically, as $\mathcal R_e\to \infty$, the velocity field will develop infinite gradients, and becomes \textit{rough} \cite{Kol41,Fri95}.

Rephrased in terms of regularity of functions, this phenomenology, mainly due to Kolmogorov \cite{Kol41,Fri95}, and that we depicted schematically in the former paragraph, has a precise formulation if we assume underlying Gaussian statistics: the velocity field $\textbf{u}(\textbf{x},t)$ is a finite-variance (zero-average under the assumptions of homogeneity and isotropy) Gaussian random field, which spatial increments over $\ell$ along any directions behave as those of a fractional Brownian motion \cite{ManVan68} of parameter $H=1/3$ (see Refs \cite{RobVar08,CheRob10,PerGar16} for recent developments on this matter).

Until now, we focused on the statistical characterization of the scale-invariance properties of the velocity field in \textit{space}. We would like now to focus on the stochastic modeling of the \textit{temporal} structure of turbulence. A similar phenomenology can be developed for the Lagrangian velocity $\textbf{v}(t)=\textbf{u}(\textbf{r}(t),t)$ of a fluid particle, initially at the position say $\textbf{r}_0$, along its trajectory $\textbf{r}(t)$ defined through the Eulerian velocity field $\textbf{u}$ as $\partial \textbf{r}(t)/\partial t = \textbf{u}(\textbf{r}(t),t)$. In this case, experimental and numerical observations (see the reviews \cite{Yeu02,TosBod09,PinSaw12} and references therein)  suggest strongly, as expected from a dimensional analysis, that Lagrangian velocity $v$ is also a finite-variance process with the same standard deviation $\sigma$, typically correlated over the large time scale $T=L/\sigma$, and which increments behave, up to the variance, at small scale as those of the Brownian motion. In a Gaussian approximation, which is a good starting point in the context of stochastic modeling, but which is too simplistic to reproduce all the observed statistics (in particular the intermittency phenomenon \cite{Yeu02,TosBod09,PinSaw12,CheCas12}), a natural stochastic model for the dynamics of $v$ would be a Ornstein-Uhlenbeck process that reads
\begin{equation}\label{eq:OU}
dv(t)= -\frac{1}{T}v(t)dt+\sigma\sqrt{\frac{2}{T}} dW(t),
\end{equation}
where $W$ is a Wiener process. The statistical properties of the solution $v$ to this stochastic differential equation (Eq. \ref{eq:OU}) are well known, and reproduce adequately at this level of phenomenology the fluctuations of the turbulent Lagrangian velocity. In this view, Lagrangian velocity is  H\"{o}lder continuous of parameter $H=1/2$.

If now we are interested in the temporal description of the fluctuations of the full Eulerian velocity field $\textbf{u}(x,y,z,t)$, or let us say at the description at a fixed position of one component of the velocity field, for example $u(t) \equiv u_x(x_0,y_0,z_0,t)$, then dimensional arguments \cite{TenLum72} ask for a finite variance process which is H\"{o}lder continuous as those of a fractional Brownian motion of parameter $H=1/3$, thus less regular than the Lagrangian counterpart $v$. This extrapolated non-differential behavior at infinite Reynolds number can be observed in numerical simulations, as it has been done in Ref. \cite{CheRou05}, and recently revisited in the context of renormalization group theory \cite{CanRos17}. In this case, a natural stochastic model for the dynamics of $u$ would be given by a generalization of the Ornstein-Uhlenbeck process (Eq. \ref{eq:OU}) of the form
\begin{equation}\label{eq:FOUIntro}
du(t)= -\frac{1}{T}u(t)dt+`` dW_H(t)",
\end{equation}
that would lead to a stationary regime, in which $u$ is of finite variance, and behaves at small scales as a fractional Brownian motion of parameter $H\in]0,1[$. All the purpose of this article is devoted to give a precise mathematical (and numerical) meaning to the random measure $``dW_H"$ that enters the dynamics of $u$ (Eq. \ref{eq:FOUIntro}). Let us mention that this level of roughness ($H<1/2$) is also a hallmark of subdiffusive dynamics, as it is studied in \cite{MagWer09}.

A natural way to generalize the dynamics of the Ornstein-Uhlenbeck process (Eq. \ref{eq:OU}) to the fractional case, i.e. a fractional Ornstein-Uhlenbeck process (Eq. \ref{eq:FOUIntro}), that would lead to a stationary finite variance process with a appropriate rough behavior at small scales, is to consider a fractional Brownian motion (fBm) $W_H(t)$ of parameter $H$. The fBm is a well defined probabilistic object \cite{ManVan68}, and allows to define accordingly the integrated dynamics as $u(t)-u(0)= -\frac{1}{T}\int_0^tu(s)ds+W_H(t)-W_H(0)$, as it is studied in \cite{CheKaw03,KaaSal11,HuNua10}. This procedure is clearly well defined, can be extended to a more general framework allowing to elaborate a stochastic calculus with respect to fractional Brownian motion \cite{PipTaq00,BiaHu08,Mis08}. Such a dynamics for $u$ indeed leads to the statistical properties that have been listed. But it eludes the question regarding the meaning of the infinitesimal increment $``dW_H(t)"$.

Nonetheless, as already proposed in \cite{ManVan68}, it is tempting to use the so-called fractional Gaussian noise, that we will specify later, to give a meaning to this infinitesimal increment.  Indeed, for $H>1/2$, this fractional Gaussian noise has a well-behaved covariance, which is bounded for non vanishing argument. The purpose of this article is to include in this picture the (very) rough case $H<1/2$, as demanded by the physics of turbulence ($H=1/3$), that requires a different method of construction due to the pathological nature of the fractional Gaussian noise (its covariance is no more bounded) at these levels of roughness. This will be achieved using a regularized form of this noise over a small time scale, say $\epsilon$, supplemented by a Gaussian white noise weighted by a factor that diverges with $\epsilon$ (see the following Eq. \ref{eq:FirstDefDW}). As we will see, an additional correlation between these two plays a key role. Once injected in the dynamics (Eq. \ref{eq:FOUIntro}), we will then study the limiting behavior of the marginals of the Gaussian process $u$ when $\epsilon \to 0$. In this sense, the regularization procedure brings to light the underlying mechanisms at play, and allows to give a precise mathematical meaning (up to $\epsilon>0$) to the random measure $dW_H(t)$ that enters in the dynamics of $u$ (Eq. \ref{eq:FOUIntro}). This overall picture clarifies that the expected dynamics leading to stationary and rough processes has to be non Markovian. In particular, we recover standard interpretations pertaining to fractional Brownian motions, namely that the infinitesimal increment is positively correlated (or persistent) for $H>1/2$, and negatively correlated (or anti-persistent) for $H<1/2$. Incidentally, it also gives a way to build up a realistic numerical approximation of the trajectories of $u$ at a given regularization scale $\epsilon$. Finally, this approach allows to deal with the boundary value $H=0$ in a non ambiguous manner, a case that is tricky to consider using the standard approach consisting in working directly with a fractional Brownian motion, as it is proposed in Refs. \cite{CheKaw03,KaaSal11,HuNua10,PipTaq00,BiaHu08,Mis08}.

Going back to the physics of turbulence, let us mention that the fractional Ornstein-Uhlenbeck process as a model of the temporal structure of the velocity field has been already proposed and studied in the literature \cite{Sha95,Sch12} in the usual sense developed in the mathematical references \cite{CheKaw03,KaaSal11,HuNua10,PipTaq00,BiaHu08,Mis08}. The novelty of the present work, besides the theoretical and numerical aspects mentioned earlier, is the introduction of a new characteristic time scale $\epsilon$ that eventually depends on viscosity, or equivalently on the Reynolds number $\mathcal R_e$. Following dimensional arguments developed in \cite{TenLum72}, we expect $\epsilon$ to coincide with the Kolmogorov time scale $\propto T\mathcal R_e^{-1/2}$ in a Lagrangian context (Eq. \ref{eq:OU}) and with the sweeping time scale $\propto T\mathcal R_e^{-3/4}$ in the Eulerian reference frame (Eq. \ref{eq:FOUIntro}). From a physical point of view, in both cases, it is expected that temporal velocity profiles are smooth below $\epsilon$. In this article, we will mainly focus on the limit $\epsilon \to 0$, but a further modeling step consisting in filtering the Gaussian white noise entering in the construction over the time scale $\epsilon$ could be performed to impose this smooth behavior imposed by viscosity. We keep this aspect for future investigations.

We give in Section \ref{Sec:SetNot} a proper meaning of this stochastic differential equation (Eq. \ref{eq:FOUIntro}) and define the random measure $dW_H$ that enters in the dynamics, more precisely a regularized version of it $dW_{\epsilon,H}$ over a small time scale $\epsilon$, and set our notations. The statistical properties in the stationary regime and their limit when $\epsilon\to 0$ of the unique solution of such a differential equation are exposed in Section \ref{Sec:PresRes}. Section \ref{Sec:GenFrame} is devoted to the proofs of the propositions made in Section \ref{Sec:PresRes} in a general framework. We gather all the numerical experiments in Section \ref{Sec:Numerics} and conclude our work in Section \ref{Sec:Conclu}.

\section{Setup and notations}\label{Sec:SetNot}

We are interested here in studying the statistical properties of the solution $X_{\epsilon,H}(t)$ of the following linear stochastic differential equation
\begin{equation}\label{eq:sdeX}
dX_{\epsilon,H}(t) = -\alpha X_{\epsilon,H}(t)dt+dW_{\epsilon,H}(t),
\end{equation}
where the random measure $dW_{\epsilon,H}$ is defined by
\begin{align}\label{eq:FirstDefDW}
dW_{\epsilon,H}(t)=\omega_{\epsilon,H}(t)dt + \epsilon^{H-\frac{1}{2}}dW(t),
\end{align}
with $H\in ]0,1[$, $\alpha>0$ the inverse of a characteristic time scale, $W$ an instance of the Wiener process, and 
\begin{equation}\label{eq:omega}
\omega_{\epsilon,H}(t) = \left(H-\frac{1}{2}\right) \int_{-\infty}^t \frac{1}{(t-s+{\epsilon})^{\frac{3}{2}-H}} dW(s),
\end{equation}
a regularized version of the fractional Gaussian noise \cite{ManVan68} over the time scale $\epsilon>0$. Remark that the noise $\omega_{\epsilon,H}$ (Eq. \ref{eq:omega}) is a zero-average and finite-variance Gaussian stationary process for any $\epsilon>0$. Remark also that the very same instance of the Wiener process $W$ enters in both the dynamics (Eq. \ref{eq:sdeX}) and in the definition of $\omega_{\epsilon,H}$ (Eq. \ref{eq:omega}), making these two correlated.

The unique solution $X_{\epsilon,H}(t)$ of Eq. \ref{eq:sdeX} with initial condition, for instance, $X_{\epsilon,H}(t_0)=0$ can be conveniently written as
\begin{equation}\label{eq:X}
X_{\epsilon,H}(t) = X_{\epsilon,H}^{(1)}(t)+X_{\epsilon,H}^{(2)}(t),
\end{equation}
with
\begin{equation}\label{eq:X1}
X_{\epsilon,H}^{(1)}(t) = \int_{t_0}^t e^{-\alpha (t-s)}\omega_{\epsilon,H}(s)ds 
\end{equation}
which coincides with the fractional Ornstein-Uhlenbeck (FOU) process \cite{CheKaw03,KaaSal11,HuNua10} in the asymptotic limit $\epsilon\to 0$ for $H>1/2$, and
\begin{equation}\label{eq:X2}
X_{\epsilon,H}^{(2)}(t) =\epsilon^{H-\frac{1}{2}}\int_{t_0}^t e^{-\alpha (t-s)}dW(s)
\end{equation}
a standard Ornstein-Uhlenbeck (OU) process. The very origin of taking the sum of these two correlated Gaussian processes will become clear later when we will present a heuristics, proposed in Ref. \cite{Sch03}, that we adapt to our context. Let us now look at the statistical properties of the Gaussian process $X_{\epsilon,H}(t)$ in the stationary regime, if any, and asymptotically when the regularizing scale $\epsilon$ tends to zero.

\section{Convergence of the covariance in the stationary regime as the regularization scale $\epsilon$ goes to zero}\label{Sec:PresRes}
\subsection{Statement of the results}\label{chap:State}
For $\alpha>0$ and a given $H\in]0,1[$, the Gaussian process $X_{\epsilon,H}(t)$ (Eq. \ref{eq:X}) reaches a stationary regime. Furthermore, it is zero-average and the variance remains bounded when $\epsilon\to 0$. We note
\begin{align}\label{eq:PredVarianceAsympt}
\E X_H^2 &=  \lim_{\epsilon\to 0}\lim_{t\to \infty}\E \left[\left(X_{\epsilon,H}(t)\right)^2\right]\notag\\
&=\frac{\alpha^{-2H}\left[\Gamma\left( H+\frac{1}{2}\right)\right]^2 }{2\sin(\pi H)}<\infty,
\end{align}  
where enters the Gamma function $\Gamma(z)=\int_0^{\infty}x^{z-1}e^{-x}dx$ defined $\forall z>0$. Let us call then $\delta_\tau X_{H}$ the corresponding increment over $\tau$, note its variance as 
$$\E \left(\delta_\tau X_{H}\right)^2= \lim_{\epsilon\to 0}\lim_{t\to \infty}\E \left[\left(X_{\epsilon,H}(t+\tau)-X_{\epsilon,H}(t)\right)^2\right].$$
For $H\in]0,1[$, we have the following behavior of the increment variance at small scales
\begin{align}\label{eq:Struct2Asympt}
\E \left(\delta_\tau X_{H}\right)^2\build{\sim}_{\tau\to 0}^{}\frac{1}{\sin(\pi H)} \frac{\left[\Gamma\left( H+\frac{1}{2}\right)\right]^2 }{\Gamma(2H+1)}|\tau|^{2H}.
\end{align}

\subsection{Remarks}\label{Sec:remarksTheo}

Given the statements of Section \ref{chap:State}, we can see that the proposed dynamics (Eq. \ref{eq:sdeX}), for a given $\epsilon>0$ and $0<H<1$, converges when $t\to\infty$ towards a stationary process $X_H$ which variance remains bounded. Its precise value can be calculated (see the section devoted to proofs) and is strictly positive. Furthermore, the increments of this process $X_H$ behave as those of the fractional Brownian motion (Eq. \ref{eq:Struct2Asympt}). 

\subsubsection{The case $H=1/2$}

The case $H=1/2$ corresponds to a standard Ornstein-Uhlenbeck process $X_{\epsilon,1/2}=X_{1/2}$ since $\omega_{\epsilon,1/2}=0$ (Eq. \ref{eq:omega}) at any time, of variance $1/(2\alpha)$ and increments behaving as in Eq. \ref{eq:Struct2Asympt} (for $H=1/2$).

\subsubsection{The case $H\in]1/2,1[$}

The case $H\in]1/2,1[$ has already been studied in Ref. \cite{CheKaw03} and can be understood easily. For $H>1/2$, the increment of the Wiener process entering in the dynamics (i.e. the third term in the RHS of Eq. \ref{eq:sdeX}) will have no contribution when $\epsilon\to 0$ since it is multiplied by a factor $\epsilon^{H-1/2}$. In this case, both variance and increment variance of $X_{\epsilon,H}$ are given by those of $X_{\epsilon,H}^{(1)}$ (Eq. \ref{eq:X1}) that can be shown to remain bounded when $\epsilon\to 0$ with a proper scaling (Eq. \ref{eq:Struct2Asympt}) at small scales. The proof relies on the fact that the noise  $\omega_{\epsilon,H}$ (Eq. \ref{eq:omega}) entering in the dynamics, even if its variance diverges when $\epsilon\to 0$ (as expected from a fractional Gaussian noise),  has a bounded covariance structure in this limit, such that the variance of $X_{\epsilon,H}^{(1)}$ remains also bounded. We develop these ideas in Section \ref{Sec:HSUP12}.

\subsubsection{The case $H\in]0,1/2[$}

The case $H\in]0,1/2[$ is more surprising since both variances of $X_{\epsilon,H}^{(1)}$ (Eq. \ref{eq:X1}) and $X_{\epsilon,H}^{(2)}$ (Eq. \ref{eq:X2}) diverge when $\epsilon\to 0$ whereas the variance of $X_{\epsilon,H}=X_{\epsilon,H}^{(1)}+X_{\epsilon,H}^{(2)}$ will remain bounded in this limit. Cancelations in the variance will take place because of the negative correlation existing in between the processes $X_{\epsilon,H}^{(1)}$ and $X_{\epsilon,H}^{(2)}$. This negative correlation originates from the fact that the Gaussian noise $\omega_{\epsilon,H}$ (Eq. \ref{eq:omega}) is made up of the very same instance of the Wiener process $W$ that enters in the dynamics of $X_{\epsilon,H}$ (Eq. \ref{eq:sdeX}).

To justify the form of the infinitesimal increment $dW_{\epsilon,H}$ (Eq. \ref{eq:FirstDefDW}) and make a connection with fBm \cite{ManVan68}, let us here rephrase some arguments developed in  \cite{Sch03}.  To do so, consider a regularized version $W_{\epsilon,H}(t)$, over $\epsilon>0$, of a fBm of parameter $H$, that is
\begin{align}\label{eq:RegulfBm}
W_{\epsilon,H}(t)&-W_{\epsilon,H}(0) = \notag \\
&\int_{-\infty}^0 \left[(t-s+\epsilon)^{H-1/2}-(-s+\epsilon)^{H-1/2}\right]dW(s)\notag\\
&+\int_{0}^t (t-s+\epsilon)^{H-1/2}dW(s).
\end{align}
Remark that the regularization procedure entering in Eq. \ref{eq:RegulfBm} is not unique.  For instance, we could also define a regularized version of the fBm as its convolution with a mollifier (properly rescaled over $\epsilon$) as it is done in \cite{ManVan68}. In all cases, these regularized versions (including Eq. \ref{eq:RegulfBm}) can easily be shown to converge toward the canonical fBm when $\epsilon\to 0$. The regularization procedure that we propose (Eq. \ref{eq:RegulfBm}) allows to compute in a simple way the infinitesimal increment of the process $W_{\epsilon,H}$ that will eventually coincide with the noise $dW_{\epsilon,H}$ that we defined in Eq. \ref{eq:FirstDefDW}. Indeed, regrouping terms in a convenient way, we get from Eq. \ref{eq:RegulfBm}
\begin{align}\label{eq:HeurDyn}
&dW_{\epsilon,H}(t)\equiv W_{\epsilon,H}(t+dt)-W_{\epsilon,H}(t) = \notag \\
&\int_{-\infty}^t\left[(t+dt-s+\epsilon)^{H-1/2}-(t-s+\epsilon)^{H-1/2}\right]dW(s)\notag\\
&+\int_t^{t+dt}(t+dt-s+\epsilon)^{H-1/2}dW(s).
\end{align}
Performing then a Taylor development (as $dt\to 0$) inside the first integral entering in the RHS of Eq. \ref{eq:HeurDyn}, we recover the contribution of order $dt$, proportional to $\omega_{\epsilon,H}$ (Eq. \ref{eq:omega}) entering in our initial proposition for the noise $dW_{\epsilon,H}$ (Eq. \ref{eq:FirstDefDW}). The second integral entering in the RHS of Eq. \ref{eq:HeurDyn} justifies the second term $\epsilon^{H-1/2}dW(t)$ entering in Eq. \ref{eq:FirstDefDW}. This is indeed true in average and for the variance. More mathematical developments would be needed to fully justify this locally (pathwise). Nonetheless, we can see that the proposition that we made for the noise $dW_{\epsilon,H}$ (Eq. \ref{eq:FirstDefDW}) can be justified in a convincing, but not fully rigorous, manner while considering the infinitesimal increment of a regularized version of the fBm (Eq. \ref{eq:RegulfBm}).

Let us also mention recent works \cite{FyoKho16} that use a similar, although different, type of regularization procedure to define and use a fractional Gaussian noise for the very rough case $H<1/2$ (see also Ref. \cite{Unt09} for similar developments). Their process (see Eq. 1.9 of Ref.  \cite{FyoKho16}) shares similar features as the fractional Brownian motion of Ref. \cite{ManVan68} when their regularization parameter tends to zero. More work is needed to characterize precisely the differences between the construction of Ref.  \cite{FyoKho16} and ours, and this is beyond the scope of the present article.

\subsubsection{The case $H=0$}\label{Sec:RemH0}

The case $H=0$ is of special interest and is fully treated in Ref. \cite{PerMor17} in view of applications in theory of fluid turbulence. It can be shown, and this is not purpose of the article (see Ref. \cite{PerMor17} for developments on this matter), that indeed the process $ X_{\epsilon,0}(t)$ reaches a stationary regime, but the variance diverges when $\epsilon\to 0$ as
$$\lim_{t\to \infty} \E \left[ X_{\epsilon,0}^2 (t) \right]\build{\sim}_{\epsilon\to 0}^{}\ln\frac{1}{\epsilon},$$  
and, furthermore, for $\tau>0$,
$$ \lim_{\epsilon\to 0}\lim_{t\to \infty} \E \left[ X_{\epsilon,0}(t) X_{\epsilon,0}(t+\tau) \right]  \build{\sim}_{\tau\to 0}^{}\ln\frac{1}{\tau}.$$  
As we can see, $X_{\epsilon,0}$ converges towards a Gaussian process which is logarithmically correlated in time, and thus of infinite variance. This type of random distributions have been studied for some time (see the review Ref. \cite{DupRho14} and citations therein), find many applications in the theory of multiplicative chaos \cite{RhoVar14}, which is used as a model of the intermittency phenomenon in turbulence (see \cite{Fri95,PerGar16}), explaining why it was considered in Refs. \cite{Sch03,PerMor17}. Once again, the construction of Ref.  \cite{FyoKho16} is similar to the one of the present work, although different, and more work is needed to compare them precisely.

\section{A more general framework and proofs}\label{Sec:GenFrame}
\subsection{Stochastic integration against fractional Gaussian noise}

In order to show the statistical properties of the Gaussian process $X_{\epsilon,H}$ in the limit $\epsilon\to 0$ as announced in Section \ref{chap:State} (Eqs.  \ref{eq:PredVarianceAsympt} and \ref{eq:Struct2Asympt}), let us rephrase former considerations in a more general way. We are here interested in calculating the covariance of the Gaussian processes obtained as a linear operation on the Gaussian random measure $dW_{\epsilon,H}(t)$ (Eq. \ref{eq:FirstDefDW}) that we recall here for convenience
\begin{align}\label{eq:dWH}
dW_{\epsilon,H}(t) = \omega_{\epsilon,H}(t)dt+\epsilon^{H-1/2}dW(t),
\end{align}
 where $W(t)$ is a instance of the Wiener process over $t\in\mathbb R$,   and  $\omega_{\epsilon,H}(t)$ the respective causal fractional Gaussian that we consider in Eq. \ref{eq:omega}. Henceforth, we will only consider convolutions as linear operations, and so only consider stationary processes (of possibly infinite variance), for the sake of simplicity. Following developments could be adapted to non stationary processes, we keep them for future investigations. As far as we are concerned, in particular while being interested by the statistics of the fractional Ornstein-Uhlenbeck process $X_{\epsilon,H}(t)$ (Eq. \ref{eq:X}), we can choose as a initial condition $X_{\epsilon,H}(t_0=-\infty)=0$, such as $X_{\epsilon,H}(t)$ is directly stationary. With these given precisions, for any suitable test functions $f$ and $g$ that we will specify latterly, let us define the covariance function
\begin{align}\label{eq:CovGenH}
&\mathcal C_{\epsilon,H}^{f,g}(t_2-t_1)\notag\\
&=\E \left[\int_{\mathbb R} f(t_1-s_1)dW_{\epsilon,H}(s_1)\int_{\mathbb R} g(t_2-s_2)dW_{\epsilon,H}(s_2)\right]\notag\\
&=\int_{\mathbb R^2} f(t_1-s_1)g(t_2-s_2)\E \Big[dW_{\epsilon,H}(s_1)dW_{\epsilon,H}(s_2)\Big].
\end{align}
Let us then rephrase in this more general framework the results of Section \ref{chap:State}.

We have, for any $H\in]0,1[$ and a appropriate set of test function $f$ and $g$, the following explicit expression of the correlation function $\mathcal C_{\epsilon,H}^{f,g}(\tau)$ when the regularization scale $\epsilon$ tends to 0:
\begin{widetext}
\begin{equation}\label{eq:CovGenHAsympt}
\mathcal C_{H}^{f,g}(\tau)=\lim_{\epsilon\to 0}\mathcal C_{\epsilon,H}^{f,g}(\tau)=-\frac{1}{2}\frac{1}{\sin(\pi H)}\frac{\left[\Gamma\left( H+\frac{1}{2}\right)\right]^2 }{\Gamma(2H)}\int_0^{+\infty}\Big[ (f\star g)'(\tau+h)-(f\star g)'(\tau-h)\Big]h^{2H-1}dh,
\end{equation}
\end{widetext}
where the symbol $\star$ stands for the correlation product, i.e.
\begin{align}\label{eq:StarProd}
(f\star g)(h)=\int_{\mathbb R} f(s)g(h+s)ds.
\end{align}
and where $'$ stands for the derivative. As we will see more precisely in the following, the correlation function $\mathcal C_{H}^{f,g}$ (Eq. \ref{eq:CovGenHAsympt}) makes sense for any $H\in]0,1[$ since the function $h^{2H-1}$ is locally integrable everywhere. As far as the test functions $f$ and $g$ are concerned, we are asking them to be such that the derivatives of their correlation product decreases fast enough at large arguments such that the integral entering in Eq. \ref{eq:CovGenHAsympt} exists. More precise constraints on test functions in terms of functional spaces are developed in particular in Refs. \cite{PipTaq00,BiaHu08,Mis08,Jol07}. We will see in Section \ref{Sec:AppFOUTheo} how to apply Eq. \ref{eq:CovGenHAsympt} to the case of the fractional Ornstein-Uhlenbeck $X_{\epsilon,H}(t)$ in order to prove the results of Section \ref{chap:State}. Let us discuss now and demonstrate the behavior of the correlation function  $\mathcal C_{H}^{f,g}$ (Eq. \ref{eq:CovGenHAsympt}) given a value of the parameter $H$, in the spirit of the remarks made in paragraph \ref{Sec:remarksTheo}.

\subsection{Dependence of the covariance on $H$}

\subsubsection{Integration over the Wiener process}
The case $H=1/2$ corresponds to the integration over the Wiener process, known as the It\^{o} integral \cite{Nua00}. In this case, the random measure $dW_{\epsilon,1/2}=dW$ (Eq. \ref{eq:dWH}) is the increment of the Wiener process, and is independent on $\epsilon$, and we get the following simple expression for the covariance function (Eq. \ref{eq:CovGenH})
\begin{align}\label{eq:CovGen12}
\mathcal C_{\epsilon,1/2}^{f,g}(\tau)= (f\star g)(\tau).
\end{align}
Remark that in this case, the expression of the correlation function Eq. \ref{eq:CovGen12} corresponds to the more general expression given in Eq. \ref{eq:CovGenHAsympt} when $H=1/2$.

\subsubsection{Integration over the fractional Gaussian noise when $H>1/2$}\label{Sec:HSUP12}
The case $H>1/2$ is also well understood in this framework \cite{PipTaq00,Jol07}. For this range of $H$, we can give a meaning of the limiting value of the correlation function $\mathcal C_{\epsilon,H}^{f,g}(\tau)$ (Eq. \ref{eq:CovGenH}) when $\epsilon\to 0$. In other words, besides giving a way to perform numerical simulations of the random measure 
$dW_{\epsilon,H}(t)$ (Eq. \ref{eq:dWH}), as we will see in Section \ref{Sec:Numerics}, there is no need theoretically to introduce a regularization at the scale $\epsilon$ in the construction, and we can safely take the limit $\epsilon\to 0$ pointwise (see for instance Refs.  \cite{PipTaq00,Jol07} for mathematical developments in the framework of random distributions). Taking a pointwise limit, we have formally
\begin{align}\label{eq:CovMeasHge2}
&\lim_{\epsilon\to 0}\E \Big[dW_{\epsilon,H}(s_1)dW_{\epsilon,H}(s_2)\Big]\notag \\
&=\lim_{\epsilon\to 0}\E \Big[\omega_{\epsilon,H}(s_1)\omega_{\epsilon,H}(s_2)\Big]ds_1ds_2\notag\\
&= \left(H-\frac{1}{2}\right)^2\int_{u=0}^{\infty}\frac{1}{u^{3/2-H}}\frac{1}{(u+|s_1-s_2|)^{3/2-H}}duds_1ds_2\notag\\
&= \left(H-\frac{1}{2}\right)\frac{1}{\sin(\pi H)}\frac{\left[\Gamma\left( H+\frac{1}{2}\right)\right]^2 }{\Gamma(2H)}|s_1-s_2|^{2H-2}ds_1ds_2.
\end{align}
Inserting then Eq. \ref{eq:CovMeasHge2} into the expression of the correlation function (Eq. \ref{eq:CovGenH}), making the change of variable $h=s_1-s_2$ and integrating over $s_2$, we get
\begin{align}\label{eq:CovGenHge2}
\mathcal C_{\epsilon,H>\frac{1}{2}}^{f,g}(\tau)=&\left(H-\frac{1}{2}\right)\frac{1}{\sin(\pi H)}\frac{\left[\Gamma\left( H+\frac{1}{2}\right)\right]^2 }{\Gamma(2H)}\notag\\
&\times \int_{\mathbb R}(f\star g)(\tau+h)|h|^{2H-2}dh,
\end{align}
which eventually coincides with the given general expression for this correlation function (Eq. \ref{eq:CovGenHAsympt}) after splitting the integral in two and performing a integration by parts over the dummy variable $h$. Remark that the intermediate expression of the correlation function (Eq. \ref{eq:CovGenHge2})  makes perfectly sense since the singularity $|h|^{2H-2}$ is locally integrable in the neighborhood of the origin for $1/2<H<1$.

\subsubsection{Integration over the fractional Gaussian noise when $H<1/2$}

This higher level of roughness requires some more work since the correlation function of the fractional Gaussian noise $\omega_{\epsilon,H}$ entering in Eq. \ref{eq:CovMeasHge2} has no meaning in the limit $\epsilon\to 0$. The mechanism of regularization over $\epsilon$ will here play a key role, and will allow several key cancellations of diverging quantities, such that the correlation function $\mathcal C_{\epsilon,H}^{f,g}(\tau)$ (Eq. \ref{eq:CovGenH}) remains a bounded function of $\epsilon$. Remark first that, formally, we can write
\begin{align}\label{eq:ExpDebHle12}
&\E \Big[dW_{\epsilon,H}(s_1)dW_{\epsilon,H}(s_2)\Big]=\\
&\left(\E \Big[\omega_{\epsilon,H}(s_1)\omega_{\epsilon,H}(s_2)\Big] +\epsilon^{2H-1}\delta(s_1-s_2)\right)ds_1ds_2+\notag\\
&\epsilon^{H-\frac{1}{2}}\left(\E \Big[\omega_{\epsilon,H}(s_1)dW(s_2)\Big]ds_1+\E \Big[\omega_{\epsilon,H}(s_2)dW(s_1)\Big]ds_2\right),\notag
\end{align}
where the contribution of the Wiener process  corresponding to $\E [dW(s_1)dW(s_2)]=\delta(s_1-s_2)ds_1ds_2$ can be conveniently noted with a Dirac function $\delta$. Let us work out first the contribution coming from the fractional Gaussian noise $\omega_{\epsilon,H}$. It reads
\begin{align}
&\E \Big[\omega_{\epsilon,H}(s_1)\omega_{\epsilon,H}(s_2)\Big]=\left(H-\frac{1}{2}\right)^2\notag\\
&\times\int_{u=0}^{\infty}\frac{1}{(u+\epsilon)^{3/2-H}}\frac{1}{(u+|s_1-s_2|+\epsilon)^{3/2-H}}du,
\end{align}
which is indeed a function of $s_1-s_2$, and is not bounded in $\epsilon$. To extract diverging quantities, perform a integration by parts and obtain
\begin{align}\label{eq:CovFrNoiHge12Int}
&\E \Big[\omega_{\epsilon,H}(s_1)\omega_{\epsilon,H}(s_2)\Big]=-\left(H-\frac{1}{2}\right)\frac{\epsilon^{H-1/2}}{(|s_1-s_2|+\epsilon)^{3/2-H}}\notag\\
&-\int_{u=0}^{\infty}\frac{H-\frac{1}{2}}{(u+\epsilon)^{1/2-H}}\frac{H-\frac{3}{2}}{(u+|s_1-s_2|+\epsilon)^{5/2-H}}du.
\end{align}
Noticing that we have formally,
\begin{align}\label{eq:CovFrNoiHge12IntCancel}
\E &\Big[\omega_{\epsilon,H}(s_1)dW(s_2)\Big]ds_1+\E \Big[\omega_{\epsilon,H}(s_2)dW(s_1)\Big]ds_2\notag\\
&=\left(H-\frac{1}{2}\right)\frac{\epsilon^{H-1/2}}{(|s_1-s_2|+\epsilon)^{3/2-H}}ds_1ds_2,
\end{align}
we can see that, once inserted in Eq. \ref{eq:ExpDebHle12}, the first diverging term entering in the RHS of Eq. \ref{eq:CovFrNoiHge12Int} will cancel out with the contribution of Eq. \ref{eq:CovFrNoiHge12IntCancel}. It will only remain in Eq. \ref{eq:ExpDebHle12} the second term of he RHS of Eq. \ref{eq:CovFrNoiHge12Int} and the $\epsilon^{2H-1}\delta(s_1-s_2)ds_1ds_2$ term. Once inserted in the expression of the correlation function (Eq. \ref{eq:CovGenH}), we end up with
\begin{align*}
&\mathcal C_{\epsilon,H}^{f,g}(\tau) = \epsilon^{2H-1}(f\star g)(\tau)-\left(H-\frac{1}{2}\right)\times\\
&\int_{(\mathbb R^+)^2}\frac{\left(H-\frac{3}{2}\right)\Big[(f\star g)(\tau+h)+(f\star g)(\tau-h)\Big]}{(u+\epsilon)^{1/2-H}(u+h+\epsilon)^{5/2-H}}dudh.
\end{align*}
Perform then a integration  by parts over $h$ in the remaining integral, we can see that the first term of the RHS of the former equation (of order $\epsilon^{2H-1}$) will be compensated, and we obtain
\begin{align}\label{eq:BefLimEps}
&\mathcal C_{\epsilon,H}^{f,g}(\tau) =\left(H-\frac{1}{2}\right)\times\notag\\
&\int_{(\mathbb R^+)^2}\frac{\Big[(f\star g)'(\tau+h)-(f\star g)'(\tau-h)\Big]}{(u+\epsilon)^{1/2-H}(u+h+\epsilon)^{3/2-H}}dudh\\
&\build{=}_{\epsilon\to 0}^{}\left(H-\frac{1}{2}\right)\int_{\mathbb R^+}\frac{1}{u^{1/2-H}(u+1)^{3/2-H}}du\notag\\
&\times \int_{\mathbb R^+}\Big[(f\star g)'(\tau+h)-(f\star g)'(\tau-h)\Big]h^{2H-1}dh\notag,
\end{align}
which coincides with the expression given in Eq. \ref{eq:CovGenHAsympt} once is performed the integral over $u$.

\subsection{Fractional Ornstein-Uhlenbeck processes}\label{Sec:AppFOUTheo}

As a application of the formula given in Eq. \ref{eq:CovGenHAsympt} to our center of interest, namely the fractional Ornstein-Uhlenbeck processes $X_{\epsilon,H}(t)$ (Eq. \ref{eq:X}), with initial condition $X_{\epsilon,H}(t_0=-\infty)=0$, consider the test functions
\begin{align}\label{eq:ChoixVar}
f(t)=g(t)=e^{-\alpha t}1_{t\ge 0},
\end{align}
such that indeed
\begin{align*}
X_{\epsilon,H}(t) = \int_{\mathbb R}f(t-s)dW_{\epsilon,H}(s).
\end{align*}
Making use of
\begin{align*}
(f\star f)(h)=\frac{1}{2\alpha}e^{-\alpha |h|} \mbox{ and }(f\star f)'(h)=-\frac{h}{2|h|}e^{-\alpha |h|},
\end{align*}
once inserted in Eq. \ref{eq:CovGenHAsympt}, we obtain
\begin{align*}
 \E X_H^2 &= \mathcal C_H^{f,f}(0)\\
 &=\frac{1}{2}\frac{1}{\sin(\pi H)}\frac{\left[\Gamma\left( H+\frac{1}{2}\right)\right]^2 }{\Gamma(2H)}\int_0^{+\infty}e^{-\alpha h}h^{2H-1}dh,
 \end{align*}
which shows once simplified the proposition made in Eq. \ref{eq:PredVarianceAsympt}.

Similarly, we obtain for the increment variance, assume for instance $\tau>0$,
\begin{align*}
& \E (\delta_\tau X_H)^2 = 2\left[\mathcal C_H^{f,f}(0)-\mathcal C_H^{f,f}(\tau)\right]\\
 &=\frac{1}{\sin(\pi H)}\frac{\left[\Gamma\left( H+\frac{1}{2}\right)\right]^2 }{\Gamma(2H)}\times\\
 &\int_0^{+\infty}\left[e^{-\alpha h}-\frac{1}{2}e^{-\alpha (\tau+h)}+\frac{\tau-h}{2|\tau-h|} e^{-\alpha |\tau-h|}\right]h^{2H-1}dh.
 \end{align*}
Regrouping terms in a convenient way, we obtain
\begin{align*}
 \E (\delta_\tau X_H)^2 &=\frac{1}{\sin(\pi H)}\frac{\left[\Gamma\left( H+\frac{1}{2}\right)\right]^2 }{\Gamma(2H)}\times  \\
&\Bigg[\left[1-\cosh(\alpha\tau)\right]\int_0^{+\infty}e^{-\alpha h}h^{2H-1}dh \\
&+\int_0^{\tau}\cosh\left[\alpha (\tau-h)\right]h^{2H-1}dh\Bigg].
 \end{align*}
As $\tau\to 0$, the first term decreases toward 0 as $\tau^2$, and thus will be negligible in front of the second term that will behave as $\tau^{2H}$. To see this, rescale in the second term the dummy variable $h$ by $\tau$, then take safely the limit $\tau\to 0$ inside the integral such that to get
\begin{align*}
\E (\delta_\tau X_H)^2  \build{\sim}_{\tau\to 0}^{}\frac{1}{\sin(\pi H)}\frac{\left[\Gamma\left( H+\frac{1}{2}\right)\right]^2 }{\Gamma(2H)}\tau^{2H}\int_0^{1}h^{2H-1}dh,
 \end{align*}
which coincides with the power-law announced in Eq. \ref{eq:Struct2Asympt} once simplified.

To finish this Section, let us focus on the boundary case $H=0$. This case, fully developed in Ref. \cite{PerMor17}, deserves more care since we are dealing with infinite variance processes as mentioned in Section \ref{Sec:RemH0} in the limit $\epsilon\to 0$. This logarithmic divergence with $\epsilon$ of the variance $\mathcal C_{\epsilon,0}^{f,f}(0)$ can be seen in Eq. \ref{eq:BefLimEps}. Nonetheless, as we already explained, even if the variance of such processes diverge with $\epsilon$ (logarithmically), the correlation function $\mathcal C_0^{f,f}(\tau)$ remains bounded for $\tau>0$. Using the general expression given in Eq. \ref{eq:CovGenHAsympt}, applied to the kernel $f$ of the fractional Ornstein-Uhlenbeck process (Eq. \ref{eq:ChoixVar}), we obtain formally, noticing that the prefactor $\frac{1}{2}\frac{1}{\sin(\pi H)}\frac{\left[\Gamma\left( H+\frac{1}{2}\right)\right]^2 }{\Gamma(2H)}$ tends to 1 when $H\to 0$,
\begin{align}\label{eq:PrDivCorrH0}
& \E \Big[X_0(0)X_0(\tau)\Big] = \mathcal C_0^{f,f}(\tau)\notag\\
  &=\int_0^{+\infty}\left[\frac{1}{2}e^{\alpha(\tau+h)}-\frac{\tau-h}{2|\tau-h|} e^{\alpha|\tau-h|}\right]h^{-1}dh\notag\\
 &=-e^{-\alpha\tau}\int_0^{\tau}\frac{\sinh(\alpha h)}{h}dh+\cosh(\alpha\tau)\int_{\tau}^{\infty}\frac{1}{h}e^{-\alpha h}dh.
 \end{align}
Since the first term entering in the RHS of Eq. \ref{eq:PrDivCorrH0} remains bounded when $\tau$ gets smaller and smaler, this demonstrates the logarithmic diverging behavior of the correlation  function as $\tau\to 0$, as claimed in  Section \ref{Sec:RemH0}. This can be readily seen while performing a further integration by parts over the dummy variable $h$ entering in the second term of the RHS side of the equation.

\section{Numerical simulations}\label{Sec:Numerics}
\subsection{A periodized approximation of the fractional Gaussian noise}

We here propose a numerical method that allows to estimate the trajectories of the solution $X_{\epsilon,H}(t)$ for $H\in]0,1[$ and $\epsilon>0$. To do so, we need to come up with a approximation of the fractional Gaussian noise  $\omega_{\epsilon,H}(t)$ (Eq. \ref{eq:omega}) 
entering in the dynamics (Eq. \ref{eq:sdeX}). The first idea would be to truncate, say over a large time scale $T'$, the Wiener integral entering in (Eq. \ref{eq:omega}), and define accordingly the corresponding estimator $\hat{\omega}_{\epsilon,H,n'}(t)$ as
\begin{align}\label{eq:omegahat}
&\hat{\omega}_{\epsilon,H,n'}(t_n)\notag \\ 
&= \left(H-\frac{1}{2}\right)\sum_{i=n-n'}^n   \frac{1}{((n-i)\Delta t+{\epsilon})^{\frac{3}{2}-H}} \Delta W(t_i),
\end{align}
where $\Delta t$ is the resolution time scale of this numerical problem, $n'=\floor*{T'/\Delta t}$ the integer corresponding to the large time scale $T'$ of the truncation, and $\Delta W(t_i)=\sqrt{\Delta t}\mathcal N(0,1)$ is a discrete collection of the increment at time $t_i=i\Delta t$ of the underlying wiener process, and is made up of independent zero-average Gaussian random variable of variance $\Delta t$. As we will see in the following, the regularizing time scale $\epsilon$ is chosen as a multiple of the resolution time scale, typically $\epsilon = 10\Delta t$. The large-scale truncation $T'$ should be chosen, depending of the values of $\Delta t$, $\epsilon$ and $H$, large ``enough". Recall that the fractional noise  $\omega_{\epsilon,H}(t)$ (Eq. \ref{eq:omega}) is a well defined random process for a finite $\epsilon>0$, so its estimator $\hat{\omega}_{\epsilon,H,n'}$ (Eq. \ref{eq:omegahat}) should become independent on $n'$ as $n'\to \infty$. The reason that explains this independence on $T'$ is connected to the fact that the kernel $(t+\epsilon)^{H-3/2}$ decreases fast enough, such that its square (entering in the variance of $\omega_{\epsilon,H}$) is integrable at $t\to\infty$. Remark also that the estimator $\hat{\omega}_{\epsilon,H,n'}$ (Eq. \ref{eq:omegahat}) requires of the order $n'$ operations at each time step, which is numerically demanding. This numerical strategy has been nonetheless followed in Ref. \cite{PerMor17} in order to simulate and explore a more complex process involving a tensorial and non Gaussian generalization of the noise  $\omega_{\epsilon,H}(t)$.

Thus, instead of using Eq. \ref{eq:omegahat}, in order to minimize the error made while truncating the integral entering in Eq. \ref{eq:omega} and perform numerical simulations in a efficient way, we will rely on the discrete Fourier transform and approximate $\omega_{\epsilon,H}(t)$ (Eq. \ref{eq:omega}) by a periodical estimator  $\tilde{\omega}_{\epsilon,H}$. For full benefit of the fast Fourier transform algorithm, we consider $N=2^k$ with $k\in \mathbb N^*$. More precisely, call $N$ the number of collocation points of your numerical approximation $\tilde{\omega}_{\epsilon,H}$ and set $T_0$ the physical time duration of the trajectory, such that $\Delta t= T_0/N$. Define $t_n = n\Delta t$ for $n\in[0,N-1]$ and $t_n^{(N)}$ its periodized version, i.e. $t_n^{(N)}=t_n$ for $n\le N/2$ and $t_n^{(N)}=t_n-T_0$ for $N/2+1\le n\le N-1$ . Define then the regularized over $\epsilon$, periodized and causal kernel $\varphi(t_n)=(H-1/2)(t^{(N)}_n+\epsilon)^{H-3/2}1_{t_n^{(N)}\ge 0}$. Note by $\mathcal F$ the discrete Fourier transform. We thus get a periodized approximation $\tilde{\omega}_{\epsilon,H}$ of the fractional Gaussian noise $\omega_{\epsilon,H}(t)$ (Eq. \ref{eq:omega}) taking
\begin{equation}\label{eq:omegatilde}
\tilde{\omega}_{\epsilon,H}(t_n) = \mathcal F^{-1}\left(  \mathcal F [\varphi(t_n)]\mathcal F [\Delta W(t_n)]\right),
\end{equation}
where again, $\Delta W(t_n)$ are $N$ independent realizations of a zero-average normal random variable of variance $\Delta t$. Trajectories of $X_{\epsilon,H}(t)$ are finally obtained while integrating, using a Euler discretization scheme, their dynamics (Eq. \ref{eq:sdeX}) as, starting for example with $X_{\epsilon,H}(0)=0$, 
\begin{align}\label{eq:sdeXApprox}
X_{\epsilon,H}&(t_{n+1}) =X_{\epsilon,H}(t_{n})\notag \\
&+\left[ -\alpha X_{\epsilon,H}(t_{n}) +\tilde{\omega}_{\epsilon,H}(t_n) \right]\Delta t +\epsilon^{H-1/2}\Delta W (t_n).
\end{align}
We recall here that  the very same instance of the white noise $\Delta W$ enters both at the level of the stochastic differential equation (Eq. \ref{eq:sdeXApprox}) and in the definition of the noise $\tilde{\omega}_{\epsilon,H}$ (Eq. \ref{eq:omegatilde}). The additional implied correlation between $X_{\epsilon,H}$ and $\Delta W$ is crucial and plays a key role in the statistical properties of $X_{\epsilon,H}$ in the stationary regime.

\begin{figure}
\begin{center}
\epsfig{file=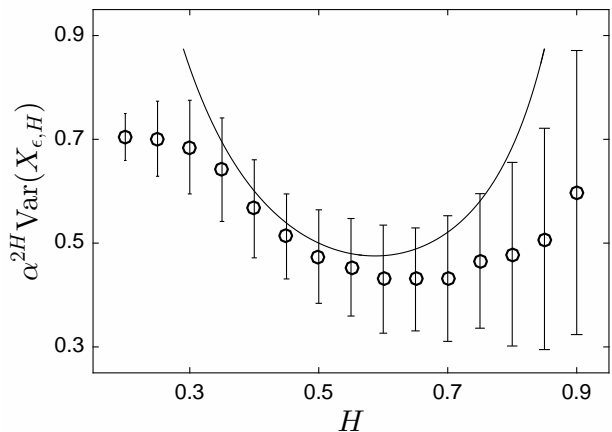,width=8cm}%8.5cm}
\end{center}
\caption{\label{fig:VarProc} Variance of the simulated trajectories of the process $X_{\epsilon,H}$ obtained while integrating the dynamics proposed in Eq. \ref{eq:sdeX} (see Eq. \ref{eq:sdeXApprox} for a discrete version). We have used  the set of parameters $T_0=1$, $N=2^{28}$, $\Delta t =T_0/N$, $\alpha=50/T_0$, $\epsilon=10\Delta t$ and for various values of the parameter $H$. The error bars are estimated as (two times) the standard deviation of the obtained variance over the $150$ realizations of the trajectories (see Section \ref{sec:NumParam} for details). We have superimposed with a solid line the corresponding theoretical prediction (Eq. \ref{eq:PredVarianceAsympt}) obtained in the stationary regime and in the limit of vanishing $\epsilon$. }
\end{figure}

\subsection{Numerical results}\label{sec:NumParam}

We consider the simulation of trajectories $X_{\epsilon,H}(t)$ of the stochastic differential equation Eq. \ref{eq:sdeX} under the approximations developed in the former Section. We choose $T_0=1$ and time is nondimensionalized accordingly. In order to minimize any statistical effects of the transitory regime, we choose $\alpha=50/T_0$, since we expect $X_{\epsilon,H}(t)$ to be correlated over typically the time scale $1/\alpha$. Present simulations are performed using $N=2^{28}$ collocation points, such that $\Delta t =T_0/N \approx 6.10^{-8}T_0$. We use for the regularization scale the value $\epsilon=10\Delta t$, and generate and analyze the statistical properties of $150$ independent trajectories of $X_{\epsilon,H}(t)$, for various values of $H$.

\begin{figure}
\begin{center}
\epsfig{file=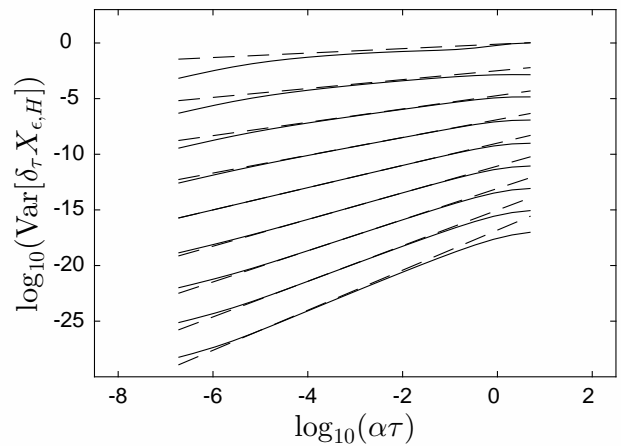,width=8cm}%8.5cm}
\end{center}
\caption{\label{fig:StructProc} Variance $\E \left(\delta_\tau X_{\epsilon,H}\right)^2= \E \left[\left(X_{\epsilon,H}(t+\tau)-X_{\epsilon,H}(t)\right)^2\right]$ of the increments as a function of the scale $\tau$ in a logarithmic fashion. We have used the same set of parameters as in Fig. \ref{fig:VarProc} (and described in Section \ref{eq:PredVarianceAsympt}). All curves are arbitraly shifted vertically for the sake of clarity. We represent nine different values of the parameter $H$, from top to bottom $H=0.1, 0.2, 0.3, 0.4, 0.5, 0.6, 0.7, 0.8$ and $0.9$. We superimpose (dashed line) with the same vertical shift our theoretical prediction pointed in Eq. \ref{eq:Struct2Asympt}. }
\end{figure}

We represent in Fig. \ref{fig:VarProc} the estimation of the variance of the simulated trajectories of the process $X_{\epsilon,H}(t)$ (Eq. \ref{eq:sdeXApprox}) as a function of the parameter $H$. We indeed observe a stationary regime, and compute the variance from $150$ independent trajectories, from which we estimate the error bars. We furthermore compare with our analytical asymptotic prediction (Eq. \ref{eq:PredVarianceAsympt}). Let us first mention that this comparison is more and more demanding as $H$ increases, since we are rescaling the estimated variance by a factor $\alpha^{-2H}$ that may become very small as $H$ approaches unity. This being said, we indeed observe that our prediction (Eq. \ref{eq:PredVarianceAsympt}) is compatible with the variance estimated on our trajectories for $0.3 \lesssim H \lesssim 0.8$. For $H\lesssim 0.3$, our prediction does not reproduce the observed variance which is characterized by a strong variability. This could be explain by several facts: (i) as a general remark, the statistical convergence of such a large-scale quantity as the variance requires many realizations and we may miss some of them, (ii)  $H\lesssim 0.3$ corresponds to the very rough case, this may require to take the small scale regularization $\epsilon$ to be taken larger than what we chose (recall that here $\epsilon=10\Delta t$), at the cost of missing the scaling properties in the asymptotics, (iii) the smoother cases $0.8\lesssim H$ have a strong statistical variability as shown by extended error bars, this might be due to a lack of statistical convergence, or a slow convergence towards the asymptotic $\epsilon\to 0$ regime. Overall, given the aforementioned limitations, our predictions seem to reproduce in a acceptable manner the variance of the simulated trajectories over a extended range of values of $H$. Let us add that numerical tests have been performed over half of the samples (i.e. 75 trajectories) without a quantitative change of the amplitudes of the error bars (data not shown), showing that discrepancies between estimated variances and theoretical prediction can be barely minimized while working on a larger set of realizations.

\begin{figure}
\begin{center}
\epsfig{file=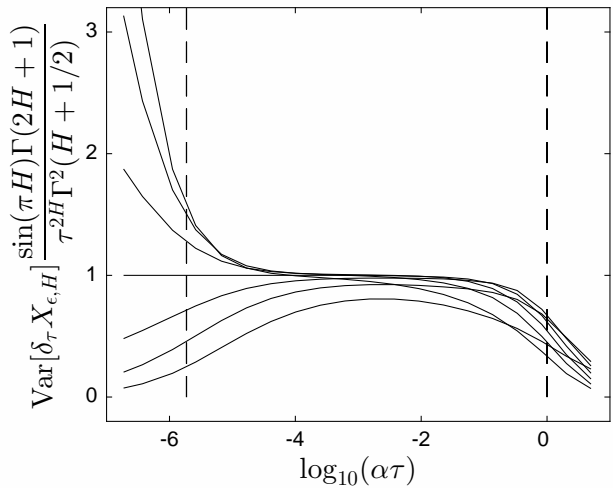,width=8cm}%8.5cm}
\end{center}
\caption{\label{fig:CompProc} Representation of the increment variance  $\E \left(\delta_\tau X_{\epsilon,H}\right)^2$ compensated by the analytical prediction pointed in Eq. \ref{eq:Struct2Asympt}, as a function of the scales $\tau$. The parameters of the simulation are the same as in Figs. \ref{fig:VarProc} and \ref{fig:StructProc}, and we represent, from top to bottom, the results for $H=0.8, 0.7, 0.6, 0.5, 0.4, 0.3, 0.2$. We furthermore superimpose with dashed-lines two characteristics time scales of the problem: the regularizing scale $\epsilon$ (left) and the large time scale $\alpha^{-1}$ (right).}
\end{figure}

We represent in Fig. \ref{fig:StructProc} the scaling behavior of the so-called structure function of second order, namely the variance $\E \left(\delta_\tau X_{\epsilon,H}\right)^2$ of the increments of the process $X_{\epsilon,H}(t)$ as a function of the scale $\tau$, for various values of the parameter $H$, from the roughest case $H=0.1$ to the smoothest case $H=0.9$. We indeed observe a power-law behavior $\tau^{2H}$ for any of the values of the parameter $H$. We superimpose on this representation the predicted behavior in the asymptotic limit $\epsilon\to 0$ (Eq. \ref{eq:Struct2Asympt}) and observe that indeed this prediction reproduces both the scale-dependence, and the prefactors. We can see also that the comparison between predictions and estimated variances deteriorates for the smallest and largest values of the parameter $H$, as it is also observed in Fig. \ref{fig:VarProc}. Since we are studying here the small scales of the process, that benefit from a large statistical sampling, we infer that we might not have reached the asymptotic regime of vanishing regularization scale $\epsilon\to 0$. 

In order to quantify precisely the differences between the observed power-laws of the increment variance and our asymptotical prediction (Eq. \ref{eq:Struct2Asympt}) as depicted in Fig. \ref{fig:StructProc}, we represent in Fig. \ref{fig:CompProc} the compensated variance $\E \left(\delta_\tau X_{\epsilon,H}\right)^2$ by our analytical prediction (Eq. \ref{eq:Struct2Asympt}), for 7 different values of the parameter $H$. For the sake of clarity, we superimpose also with vertical dashed lines the two characteristic time scales $\epsilon$ and $1/\alpha$ in between which we expect a power law behavior of exponent $2H$. We indeed observe that over  almost three decades in scale, increment variance obeys a extended power-law behavior correctly captured by our analytical prediction (Eq. \ref{eq:Struct2Asympt}) when $H>1/2$. We can also observe that our prediction deteriorates as $H$ gets smaller and smaller compared to $1/2$. Similarly, we can see two reasons for this: (i) we did not reach yet the asymptotic regime $\epsilon\to 0$ and we are observing a slow convergence towards it, and (ii) some consequences are expected in the very rough case if we chose the regularizing scale $\epsilon$ not big enough compared to the resolution scale $\Delta t$. We keep for future investigations a more developed numerical study of this process. Nonetheless, we can see that the fractional Ornstein-Uhlenbeck process can be easily simulated and our theoretical predictions (Eqs. \ref{eq:PredVarianceAsympt} and \ref{eq:Struct2Asympt}) compare reasonably well with our numerical results.

\section{Conclusions}\label{Sec:Conclu}

We have proposed a generalization (Eq. \ref{eq:sdeX}) of the Ornstein-Uhlenbeck process (Eq. \ref{eq:OU}) which is of finite-variance in the stationary regime, and which regularity is governed by the parameter $H\in]0,1[$ when the regularizing scale tends to zero, in the same fashion as the fractional Brownian motion \cite{ManVan68}. This article includes a theoretical study aimed at getting exact expressions for the variance and increment variance of the process, and their asymptotical behavior when $\epsilon\to 0$ and at vanishing scale. We propose then a numerical study showing that the fractional Gaussian noise entering in the dynamics can be approximated in a accurate way, and observe to some extend for any $H\in ]0,1[$ the convergence toward the asymptotical regime developed in the theoretical section. 

Further efforts in this spirit will be devoted to the modeling of fluid turbulence, that asks for further developments in order to give a realistic picture of the intermittency phenomenon and energy transfers in scale. We keep these perspectives for future investigations.

I would like to thank P. Abry, P. Borgnat, G. Didier, P. Flandrin, C. Garban, K. Gawedzki, L. Moriconi, R.M. Pereira, R. Rhodes, S.G. Roux and V. Vargas for many discussions concerning this matter, and J.F. Palierne for a critical proofreading of a first version of the manuscript. The author is supported by ANR grants \textsc{Liouville} ANR-15-CE40-0013 and \textsc{MultiFracs} ANR-16-CE33-0020.

%%%%%%%%%%%%

%\bibliographystyle{plain}
%\bibliographystyle{unsrt}
%\bibliography{/Users/lchevill/Dropbox/MyBibTex/mybiblioJune2015}
%\bibliography{/home/lchevill/Redac/MyBibTex/mybiblioJune2015}

\end{document}